\documentclass[twocolumn,showpacs,preprintnumbers,amsmath,amssymb]{revtex4-1}

\usepackage{graphicx}
\usepackage{dcolumn}
\usepackage{bm}
\usepackage[compact]{titlesec}

\usepackage{hyperref}
\newcommand{\half}{{{\textstyle\frac{1}{2}}}}
\newcommand{\quarter}{{{\textstyle\frac{1}{4}}}}
\newcommand{\be}{\begin{equation}}
\newcommand{\ee}{\end{equation} }
\newcommand{\beqa}{\begin{eqnarray} }
\newcommand{\eeqa}{\end{eqnarray} }
\newcommand{\ba}{\begin{array}}
\newcommand{\ea}{\end{array}}
\newcommand{\bpm}{\begin{pmatrix}}
\newcommand{\epm}{\end{pmatrix}}

\newcommand{\rmd}{{\rm d}}

\newcommand{\rmD}{{\rm D}}

\newcommand{\Apr}{{A^{\prime}{}}}
\newcommand{\Bpr}{{B^{\prime}{}}}
\newcommand{\Dpr}{{D^{\prime}{}}}

\newcommand\cHpr{\cH^{\prime}}
\newcommand\cJpr{\cJ^{\prime}}

\newcommand\hcJ{{\hat{\cal J}}}

\newcommand{\hODD}{\mathbf{O}(\hD,\hD)}
\newcommand{\ODD}{\mathbf{O}(D,D)}
\newcommand{\inODD}{\mathbf{O}(\Dpr,\Dpr)}

\newcommand\So{S_{\scriptscriptstyle{{(0)}}}}
\newcommand\cGo{\cG_{\scriptscriptstyle{{(0)}}}}
\newcommand\brcGo{\brcG_{\scriptscriptstyle{{(0)}}}}
\newcommand\hcGo{\hat{\cG}_{\scriptscriptstyle{{(0)}}}}
\newcommand\hbrcGo{\hat{\brcG}_{\scriptscriptstyle{{(0)}}}}

\newcommand{\brcG}{\bar{\cG}}

\newcommand\Tr{{\rm Tr}}

\newcommand\cA{{\cal A}}

\newcommand\cF{{\cal F}}
\newcommand\cG{{\cal G}}
\newcommand\cH{{\cal H}}

\newcommand\cJ{{\cal J}}

\newcommand\cL{{\cal L}}

\def\tx{\tilde{x}}
\def\ty{\tilde{y}}

\def\na{\nabla}

\def\breta{\bar{\eta}}

\def\brn{{\bar{n}}}
\def\brp{{\bar{p}}}
\def\brq{{\bar{q}}}
\def\brr{{\bar{r}}}

\def\brPhi{{{\bar{\Phi}}}}

\def\brP{\bar{P}}

\def\brV{\bar{V}}

\def\brcF{\bar{\cF}}

\newcommand\hp{\hat{p}}

\newcommand\hA{\hat{A}}
\newcommand\hB{\hat{B}}

\newcommand\hD{\hat{D}}

\newcommand\hV{{\hat{V}}}
\newcommand\hW{{\hat{W}}}

\newcommand\hcH{{\hat{\cH}}}

\def\dA{\dot{A}}
\def\dB{\dot{B}}
\def\dC{\dot{C}}
\def\dD{\dot{D}}
\def\dE{\dot{E}}

\def\dV{\dot{V}}

\def\dcJ{\dot{\cJ}}

\begin{document}

\title{Kaluza--Klein reduction on a maximally non-Riemannian   space is moduli-free}
\author{Kyoungho Cho}
\email{khcho23@sogang.ac.kr}
\author{Kevin Morand}
\email{morand@sogang.ac.kr}
\author{Jeong-Hyuck Park\,}
\email{park@sogang.ac.kr}
 
\affiliation{Department of Physics, Sogang University, 35 Baekbeom-ro, Mapo-gu, Seoul 04107,  Korea}

\begin{abstract}
\centering\begin{minipage}{\dimexpr\paperwidth-7.465cm}
\noindent  We propose  a novel  Kaluza--Klein  scheme  which assumes the internal space to be maximally    non-Riemannian, meaning  that  no  Riemannian  metric can be defined    for any subspace. Its description is  only possible through  Double Field Theory but  not within   supergravity.  We spell out the corresponding  Scherk--Schwarz twistable   Kaluza--Klein ansatz, and point out   that the internal space   prevents rigidly     any    graviscalar moduli.   Plugging the same ansatz into higher-dimensional pure Double Field Theory and also  to a known  doubled-yet-gauged string action, we recover   heterotic supergravity as well as   heterotic  worldsheet   action. In this way, we show that 1) supergravity  and Yang--Mills theory can be unified into higher-dimensional pure Double Field Theory, free of moduli,  and 2) heterotic string theory may have a higher-dimensional non-Riemannian origin.  

\end{minipage}
\end{abstract}

                             
\maketitle
\section{Introduction}
\vspace{5pt}
\noindent Kaluza--Klein theory attempts  to  unify General Relativity  and electromagnetism into higher-dimensional   pure gravity.  Yet, the (aesthetically unpleasing) cylindrical extra  dimension  brings along  an unwanted  additional massless scalar field,   \textit{i.e.~}radion or graviscalar  modulus, which is not observed in nature:  it would  spoil  the Equivalence Principle by appearing on the  right-hand side of the  geodesic equation.   This moduli stabilization   problem is    essentially  rooted in the fact    that  there is no natural scale in pure gravity which would  fix or stabilize  the radius of the cylinder.   The problem persists in modern  string compactifications, in view of  the arbitrary size and shape of an  internal space (mani/coni/orbifold, compact or not).  Turning on fluxes or non-perturbative  corrections might  promise to solve  the problem,  but     such  scenarios   require generically subtle analyses  regarding both the validity of the effective-field-theory approximation   and calculational control~\cite{Kachru:2003aw}  (\textit{c.f.~}\cite{Obied:2018sgi} and references therein for related   current  controversies).

  In this paper  we propose a novel  Kaluza--Klein  scheme to unify Stringy Gravity and Yang--Mills (including Maxwell), which   postulates a non-Riemannian internal space and  consequently  does not suffer from  any moduli problem.   By Stringy Gravity, we mean the string theory effective action of the  NS-NS (or  purely  bosonic) closed-string massless sector,  conventionally represented by the three fields, $\{g_{\mu\nu},B_{\mu\nu},\phi\}$. They transform into one another under  T-duality and hence form a (reducible) $\ODD$  multiplet~\cite{Buscher:1987sk,Buscher:1987qj}.  Furthermore,  within the framework of    Double Field Theory (DFT) initiated in  \cite{Siegel:1993xq,Siegel:1993th,Hull:2009mi,Hull:2009zb,Hohm:2010jy}, $\ODD$ T-duality becomes     the manifest principal  symmetry and      the effective action itself  is   to be  identified as an integral of a   stringy scalar curvature beyond Riemann. The whole  closed-string massless NS-NS sector may  then be   viewed as stringy graviton fields  consisting  of the DFT-dilaton, $d$, and the DFT-metric, $\cH_{AB}$~\cite{Angus:2018mep}. The latter satisfies   two defining properties: 
\be
\ba{ll}
\cH_{AB}=\cH_{BA}\,,\quad&\quad
\cH_{A}{}^{C}\cH_{B}{}^{D}\cJ_{CD}=\cJ_{AB}\,,
\ea
\label{defH}
\ee
\vspace{-3pt}where $\cJ_{AB}={\resizebox{.125\hsize}{!}{${\tiny\mathbf{\Big(\ba{cc} \bf{0}&\bf{1}\\\bf{1}&\bf{0}\ea\Big)}}$}}$ is the $\ODD$ invariant constant metric which can freely raise and lower the $\ODD$ vector indices, $A,B,\cdots$ (capital letters).   A pair of  symmetric  projection matrices  are then defined,
\be
\ba{ll}
P_{AB}=\half(\cJ_{AB}+\cH_{AB})\,,\quad&
\brP_{AB}=\half(\cJ_{AB}-\cH_{AB})\,,
\ea
\ee
while their  `square roots'  give twofold DFT-vielbeins, 
\be
\ba{ll}
P_{AB}=V_{A}{}^{p}V_{B}{}^{q}\eta_{pq}\,,\quad&\quad
\brP_{AB}=\brV_{A}{}^{\brp}\brV_{B}{}^{\brq}\breta_{\brp\brq}\,,
\ea
\ee
which satisfy their own  defining properties, 
\be
\ba{lll}
V_{Ap}V^{A}{}_{q}=\eta_{pq}\,,\quad&~~
\brV_{A\brp}\brV^{A}{}_{\brq}=\breta_{\brp\brq}\,,\quad&~~
V_{Ap}\brV^{A}{}_{\brq}=0\,,
\ea
\label{defV1}
\ee
or equivalently~\cite{LR}
\be
V_{A}{}^{p}V_{B}{}^{q}\eta_{pq}+\brV_{A}{}^{\brp}\brV_{B}{}^{\brq}\breta_{\brp\brq}=\cJ_{AB}\,.
\label{defV2}
\ee
Here $\eta_{pq}$ and $\breta_{\brp\brq}$ are  local  Lorentz invariant    metrics   characterizing  the twofold spin groups which  have  dimensions $\Tr(P)$ and $\Tr(\brP)$ separately.  They are distinguished by  unbarred and barred small  letter   indices.

The above stringy graviton fields constitute  the    diffeomorphic DFT-Christoffel symbols~\cite{Jeon:2010rw,Jeon:2011cn,Morand:2017fnv},
{\small{\[
\ba{ll}
\!\Gamma_{CAB}=&\!\!{2\left(P\partial_{C}P\brP\right)_{[AB]}
+2\left({{\brP}_{[A}{}^{D}{\brP}_{B]}{}^{E}}-{P_{[A}{}^{D}P_{B]}{}^{E}}\right)\partial_{D}P_{EC}}\\
{}&\!-4\!\left(\textstyle{\frac{P_{C[A}P_{B]}{}^{D}}{P_{F}{}^{F}-1}}+\textstyle{\frac{\brP_{C[A}\brP_{B]}{}^{D}}{\brP_{F}{}^{F}-1}}\right)\!\left(\partial_{D}d+(P\partial^{E}P\brP)_{[ED]}\right),
\ea
\label{Gammao}
\]}}and, consequently  with $\na_{A}:=\partial_{A}+\Gamma_{A}$,    twofold local Lorentz spin connections~\cite{Jeon:2011vx,Jeon:2011sq,Jeon:2012hp},
\be
\ba{l}
\Phi_{Apq}=V^{B}{}_{p}\na_{A}V_{Bq}=V^{B}{}_{p}\!\left(\partial_{A}V_{Bq}+\Gamma_{AB}{}^{C}V_{Cq}\right),\\
\brPhi_{A\brp\brq}=\brV^{B}{}_{\brp}\na_{A}\brV_{B\brq}=\brV^{B}{}_{\brp}\!\left(\partial_{A}\brV_{B\brq}+\Gamma_{AB}{}^{C}\brV_{C\brq}\right).
\ea
\label{PhibrPhi}
\ee
These connections form  covariant curvatures, Ricci and scalar~\cite{Jeon:2011cn,Jeon:2012hp}, of which the latter  can be constructed as
\be
\ba{l}
\So=(P^{AC}P^{BD}-\brP^{AC}\brP^{BD})S_{ABCD}\,,\\
\cGo=\cF_{ABpq}V^{Ap}V^{Bq}+\half\Phi_{Apq}\Phi^{Apq}\,,\\
\brcGo=\brcF_{AB\brp\brq}\brV^{A\brp}\brV^{B\brq}+\half\brPhi_{A\brp\brq}\brPhi^{A\brp\brq}\,,
\ea
\ee
where, with {\resizebox{.8\hsize}{!}{\,${R_{CDAB}=\partial_{A}\Gamma_{BCD}+\Gamma_{AC}{}^{E}\Gamma_{BED}-(A\leftrightarrow B)}$}}, 
\[
\ba{l}
\!S_{ABCD}:=\half\left(R_{ABCD}+R_{CDAB}-\Gamma^{E}{}_{AB}\Gamma_{ECD}\right)\,,\\
\!\cF_{ABpq}:=\na_{A}\Phi_{Bpq}-\na_{B}\Phi_{Apq}+\Phi_{Ap}{}^{r}\Phi_{Brq}-\Phi_{Bp}{}^{r}\Phi_{Arq}\,,\\
\!\brcF_{AB\brp\brq}:=\na_{A}\brPhi_{B\brp\brq}-\na_{B}\brPhi_{A\brp\brq}+\brPhi_{A\brp}{}^{\brr}\brPhi_{B\brr\brq}-\brPhi_{B\brp}{}^{\brr}\brPhi_{A\brr\brq}\,.
\ea
\]
While $\So$ coincides  exactly  with  the    well-known expression of  scalar curvature written in terms of $d, \cH_{AB}$~\cite{Hohm:2010pp}, 
\[
\ba{l}
\!\!\So=\textstyle{\frac{1}{8}}\cH^{AB}\partial_{A}\cH_{CD}\partial_{B}\cH^{CD}+\half\cH^{AB}
\partial^{C}\cH_{AD}\partial^{D}\cH_{BC}\\
\!-\partial_{A}\partial_{B}\cH^{AB}
+4\cH^{AB}(\partial_{A}\partial_{B}d-\partial_{A}d\partial_{B}d)+4\partial_{A}\cH^{AB}\partial_{B}d\,,
\ea
\]
the other two accommodate  the vielbeins and read~\cite{Cho:2015lha}
\be
\ba{rl}
+\cGo&\!\!=\half\So+2\partial_{A}\partial^{A}d-2\partial_{A}d\partial^{A}d+\half\partial_{A}V_{Bp}\partial^{A}V^{Bp}\,,\\\
-\brcGo&\!\!=\half\So-2\partial_{A}\partial^{A}d+2\partial_{A}d\partial^{A}d-\half\partial_{A}\brV_{B\brp}\partial^{A}\brV^{B\brp}\,.
\ea
\label{cGobrcGo}
\ee
Clearly their differences  would vanish upon the section condition, $\partial_{A}\partial^{A}\equiv0$, but they  provide   precisely the known `missing' pieces  in  the Scherk--Schwarz reduction of DFT  while    relaxing the  section condition  on  the internal space, as  previously  added by hand in \cite{Geissbuhler:2011mx,Aldazabal:2011nj,Grana:2012rr,Dibitetto:2012rk}.  Hence, the above vielbein formalism generates these crucial terms in a natural manner. In particular,  $\cGo\!-\brcGo$ matches  the action  adopted in  \cite{Grana:2012rr}. Below we  focus on computing   $+\cGo$ and $-\brcGo$ separately in higher dimensions   with  the Scherk--Schwarz  twisted non-Riemannian  Kaluza--Klein ansatz. \\


\section{\mbox{\!\!\!\!\!Moduli-free non-Riemannian Kaluza--Klein ansatz}}
\vspace{5pt}
\noindent With ${\hD=\Dpr+D}$, the  DFT Kaluza--Klein ansatz~\cite{Morand:2017fnv}  breaks $\hODD$ into $\inODD\times\ODD$ and takes the form:
\be
\ba{ll}
\hcH={\resizebox{.5\hsize}{!}{$\exp\!\left[\hW\right]\left(\ba{cc}\cHpr&0\\0&\cH\ea\right)\exp\!\left[\hW^{T}\right]$}}\,,
\quad&~~\hcJ={\resizebox{.17\hsize}{!}{$\left(\ba{cc}\cJpr&0\\0&\cJ\ea\right)$}}\,.
\ea
\label{ansatz}
\ee
In our notation,     hatted, primed, and unprimed symbols refer to  the    ambient, internal, and external  spaces   respectively. In particular,  the ambient doubled coordinates  split   into the internal and external ones as   
\be
\ba{ll}
z^{\hA}=(y^{A^{\prime}},x^{A})\,,\quad&\quad
\partial_{\hA}=(\partial_{A^{\prime}},\partial_{A})\,.
\ea
\label{split}
\ee
In (\ref{ansatz}),   $\hW$   is an off-block-diagonal $\mathfrak{so}(\hD,\hD)$ element which  satisfies 
\be
\hW\hcJ+\hcJ\hW^{T}=0\,,
\ee
and takes  the form,
\be
\!\hW\!=\!{\resizebox{.95\hsize}{!}{$
\left(\!\ba{cc}0&-W^{{\mathbf{c}}}\\W&\!0\ea\!\right)\!,~ 
W^{{\mathbf{c}}}_{~\Apr}{}^{\!A\!}:=W^{A}{}_{\!\Apr}= \cJ^{AB}\cJpr_{\Apr\Bpr}W_{B}{}^{\Bpr},$}}
\ee
where  the $2D\times 2\Dpr$    block,  $W_{A}{}^{\Apr}$,  should  meet~\cite{Choi:2015bga}  
\be
\ba{ll}
W_{A}{}^{\Apr}W^{A\Bpr}=0\,,\qquad&\quad W^{A}{}_{\Apr}\partial_{A}=0\,.
\ea
\label{Wconstraint} 
\ee
This    condition sets half of the components to be trivial, truncates the exponential, 
\[
\exp\!\left[\hW\right]=1+\hW+\half\hW^{2}\,,
\]
and makes  the above  ansatz    consistent with the ordinary  Kaluza--Klein  ansatz in supergravity.    Moreover, crucially for the purpose of the present work,  the   ansatz~(\ref{ansatz}) can   accommodate   non-Riemannian geometry in which the Riemannian metric cannot be defined,   see \cite{Morand:2017fnv} for classification and \cite{Lee:2013hma,Ko:2015rha,Park:2016sbw} for earlier examples  including    the attainment of  the Gomis-Ooguri non-relativistic string~\cite{Gomis:2000bd,GO}.  

 Henceforth,  we focus on a specific  internal space  of which the DFT-metric is fully $\inODD$ symmetric and   maximally  non-Riemannian, namely $(\Dpr,0)$ type as classified in   \cite{Morand:2017fnv}:
\be
\cHpr_{\Apr\Bpr}\equiv\cJpr_{\Apr\Bpr}\,.
\label{internal}
\ee     
In general,  from the defining relations~(\ref{defH}),  the infinitesimal variation of  any  DFT-metric should satisfy
\be
\delta \cHpr=P^{\prime}\delta\cHpr\brP^{\prime}+\brP^{\prime}\delta\cHpr P^{\prime}\,.
\ee
Meanwhile, the  particular   choice of the internal space~(\ref{internal}) implies ${P^{\prime}=\cJpr}$ and ${\brP^{\prime}=0}$.  Thus,  the   fluctuation must be trivial:   no graviscalar modulus  can be generated and the non-Riemannian internal space is \textit{rigid}, 
\be
\delta \cHpr_{\Apr\!\Bpr}=0\,.
\label{rigid}
\ee  
In fact,   (\ref{internal}) sets the ``twofold" internal spin group to be   $\inODD\times\mathbf{O}(0,0)$,  such that   the coset structures  of the internal and the ambient  DFT-metrics, $\cHpr$, $\hcH$, are `trivial' and `heterotic'  respectively, if $\cH$ is Riemannian~(\ref{Riemann}), 
\be
{\resizebox{.95\hsize}{!}{$
\frac{\inODD}{\inODD\times\mathbf{O}(0,0)}=\mathbf{1}\,,\quad
\frac{\hODD}{\mathbf{O}(\Dpr+1,\hD-1)\times\mathbf{O}(D-1,1)}\,.$}}
\label{coset}
\ee
The latter has  dimension ${D^{2}+2D\Dpr}$, which  matches the total degrees of the  external DFT-metric, $\cH_{AB}$  (\ref{Riemann}),  and the gravivector,  $W_{A}{}^{\Apr}$~(\ref{Wconstraint})~\cite{Quetient}, \textit{c.f.~}\cite{Maharana:1992my}.   

The corresponding DFT-vielbeins are~\cite{Morand:2017fnv}
\be
\ba{ll}
\hV_{\hA\hp}\!=\!\exp\!\left[\hW\right]\!\left(\ba{cc}
V^{\prime}_{A^{\prime}p^{\prime}}&0\\
0&V_{Ap}\ea\right),&
\hat{\brV}_{\hA\hat{\brp}}\!=\!\exp\!\left[\hW\right]\left(\ba{c}
0\\
\brV_{A\brp}\ea\right),
\ea
\label{ansatzV}
\ee
where $V^{\prime}_{A^{\prime}p^{\prime}}$ is now an invertible   `square' matrix,
 \[
V^{\prime}_{A^{\prime}p^{\prime}}V^{\prime}_{B^{\prime}}{}^{p^{\prime}}=\cJ^{\prime}_{A^{\prime}B^{\prime}}\,.
\]

\section{Reduction to heterotic DFT}
\vspace{5pt}
\noindent Before inserting  the ansatz~(\ref{ansatz}), (\ref{ansatzV}) into the ambient scalar curvatures~(\ref{cGobrcGo}), or $\hcGo,\hbrcGo$ (all hatted),  we   perform a Scherk--Schwarz twist over the internal space.  Following \cite{Cho:2015lha}, we  introduce a twisting matrix, {\resizebox{.12\hsize}{!}{${U_{\dA}{}^{A^{\prime}}(y)}$}}, which depends on the internal coordinates only and is an $\inODD$  element satisfying 
\be
U_{\dA}{}^{A^{\prime}}U_{\dB}{}^{B^{\prime}}\cJpr_{A^{\prime}B^{\prime}}=\dcJ_{\dA\dB}\,,
\ee
where $\dcJ_{\dA\dB}$ coincides with $\cJpr_{A^{\prime}B^{\prime}}$ numerically: both are $\inODD$ invariant constant metrics.  Essentially   the twist converts   all the primed indices to  dotted ones, 
\be
\ba{ll}
\! \!\!W_{A}{}^{\Apr\!}(z)=W_{A}{}^{\dA}(x)U_{\dA}{}^{A^{\prime}\!}(y)\,,&\!
V^{\prime A^{\prime}}{}_{\! p^{\prime}}(z)=\dV{}^{\dA}{}_{p^{\prime}}(x)U_{\dA}{}^{A^{\prime}\!}(y)\,,
\ea
\label{twist1}
\ee
where now  $\dV_{\dA p^{\prime}}\dV_{\dB}{}^{p^{\prime}}=\dcJ_{\dA\dB}$.    We further put
\be
\ba{lll}
\hat{d}(z)=\lambda(y)+d(x)\,,\quad&\quad
V_{Ap}(x)\,,\quad&\quad \brV_{A\brp}(x)\,,
\ea
\label{trunc}
\ee
such that the external fields are independent of the internal coordinates:  $\partial_{A^{\prime}}d=0$\,,  $\partial_{A^{\prime}}V_{Bp}=0$\,, $\partial_{A^{\prime}}\brV_{B\brp}=0$\,.

Finally, we impose the standard section condition on the external space, $\partial_{A}\partial^{A}=0$,  and the twistability conditions on the internal space separately~\cite{Grana:2012rr},\cite{Cho:2015lha}:
\be
\ba{ll}
\partial_{\Apr}\lambda-\half U^{\dA}{}_{\Apr}\partial_{\Bpr}U_{\dA}{}^{\Bpr}=0\,,\quad&\quad
f_{[\dA\dB}{}^{\dE}f_{\dC]\dD\dE}=0\,,
\ea
\label{twist2}
\ee
where  $f_{\dA\dB\dC}=3\big(\partial_{\Bpr}U_{[\dA}{}^{\Apr}\big)U_{\dB}{}^{\Bpr}U_{\dC]\Apr}$ which we require  to be  constant.  

After straightforward yet tedious computation  -- assisted by a computer algebra system~\cite{Peeters:2007wn} and through intermediate expressions like (4.14) in   \cite{Cho:2015lha} -- we obtain our main result: with the above  Scherk--Schwarz twisted Kaluza--Klein ansatz substituted,  the higher-dimensional   scalar curvatures~(\ref{cGobrcGo})  reduce precisely to (\textit{c.f.}~\cite{Jeon:2011kp,Hohm:2011ex,Grana:2012rr,Berman:2013cli,Hohm:2014sxa,Malek:2016vsh,Aldazabal:2017jhp})
\be
\ba{lll}
\!\!+2\hcGo&\!=\!&-2\hbrcGo+\textstyle{\frac{1}{3}}f_{\dA\dB\dC}f^{\dA\dB\dC}\,,\\
\!\!-2\hbrcGo&\!=\!&\So-\textstyle{\frac{1}{4}}\cH^{AC}\cH^{BD}
F_{AB}{}^{\dA}F_{CD\dA}\\
{}&{}&-\textstyle{\frac{1}{12}}\cH^{AD}\cH^{BE}\cH^{CF}\omega_{ABC}
\omega_{DEF}\\
{}&{}&+\textstyle{\frac{1}{2}}\cH^{AD}\cH^{BE}\cH^{CF}\omega_{ABC}\cH_{[D}{}^{G}\partial_{E}\cH_{F]G}\,,
\ea
\label{MAIN}
\ee
where we set Yang--Mills  and Chern--Simons terms,
\be
\ba{l}
F_{AB}{}^{\dC}=\partial_{A}W_{B}{}^{\dC}-\partial_{B}W_{A}{}^{\dC}+f_{\dA\dB}{}^{\dC}W_{A}{}^{\dA}W_{B}{}^{\dB}\,,\\
\omega_{ABC}=3W_{[A}{}^{\dA}\partial_{B}W_{C]\dA}+f_{\dA\dB\dC}W_{A}{}^{\dA}W_{B}{}^{\dB}W_{C}{}^{\dC}\,,
\ea
\label{Fomega}
\ee
of which the $\ODD$ indices are totally skew-symmetric,   $F_{AB}{}^{\dC}=F_{[AB]}{}^{\dC}$, $
\omega_{ABC}=\omega_{[ABC]}$, and further from (\ref{Wconstraint}),
\be
\ba{ll}
F^{AB}{}_{\dC}\partial_{A}=0\,,\quad&\quad\omega^{ABC}\partial_{A}=0\,.
\ea
\ee

The higher  $\hD$-dimensional  diffeomorphisms  generated by the  standard DFT  Lie derivative    give rise, with a  twisted  parameter, $\hat{\xi}{}^{\hA}=(\Lambda^{\dA}U_{\dA}{}^{\Apr},\xi^{A})$, to the $D$-dimensional diffeomorphisms plus Yang--Mills gauge symmetry, \textit{c.f.}~\cite{Grana:2012rr,Cho:2015lha}, 
\be
\ba{rll}
\delta W_{A}{}^{\dA}&=&\xi^{C}\partial_{C}W_{A}{}^{\dA}
+\left(\partial_{A}\xi^{B}-\partial^{B}\xi_{A}\right)W_{B}{}^{\dA}\\
{}&{}&+\partial_{A}\Lambda^{\dA}+f^{\dA}{}_{\dB\dC}W_{A}{}^{\dB}\Lambda^{\dC}\,,\\
\delta\cH_{AB}&=&\xi^{C}\partial_{C}\cH_{AB}+2\partial_{[A}\xi_{C]}\cH^{C}{}_{B}+
2\partial_{[B}\xi_{C]}\cH_{A}{}^{C}\\
{}&{}&+\big(W_{[A}{}^{\dA}\partial_{C]}\Lambda_{\dA}\big)\cH^{C}{}_{B}+\big(
W_{[B}{}^{\dA}\partial_{C]}\Lambda_{\dA}\big)\cH_{A}{}^{C}\,,\\
\delta d&=&\xi^{C}\partial_{C}d-\half\partial_{C}\xi^{C}\,.
\ea
\label{SYMMETRY}
\ee
In each transformation above, the first line with $\xi^{A}$ is  the (external) diffeomorphic DFT Lie derivative  and  the second   with $\Lambda^{\dA}$   is the  (internal) Yang--Mills gauge symmetry.  In fact, every single term in (\ref{MAIN}) is (external) diffeomorphism-invariant~\cite{footnoteYM}.

It is worth while  to note the only difference between    $2\hcGo$  and $-2\hbrcGo$:   the former contains  a DFT-cosmological constant term~\cite{Jeon:2011cn},    $\textstyle{\frac{1}{3}}e^{-2d}f_{\dA\dB\dC}f^{\dA\dB\dC}$, 
but the latter  does not.   As  anticipated in \cite{Hassan:1991mq,Hohm:2014sxa},  our result~(\ref{MAIN}) is manifestly symmetric for $\ODD$  as well as  any  subgroup of $\inODD$ which stabilizes the structure constant, $f_{\dA\dB\dC}$.

Finally, if we adopt the well-known Riemannian parametrization of the DFT-metric,  
\be
\ba{ll}
\cH_{AB}=\left(\ba{cc}g^{-1}&-g^{-1}B\\
Bg^{-1}&g-Bg^{-1}B\ea\right),\quad&\quad e^{-2d}=\sqrt{-g}e^{-2\phi}\,,
\ea
\label{Riemann}
\ee
and solve both the external section condition, $\partial_{A}\partial^{A}=0$, and   (\ref{Wconstraint}), by letting    $\partial_{A}=(\tilde{\partial}^{\mu},\partial_{\nu})\equiv(0,\partial_{\nu})$ and $W_{A}{}^{\dA}\equiv(0,W_{\nu}{}^{\dA})$,  our main result~(\ref{MAIN})  reproduces    the heterotic supergravity  action~\cite{Bergshoeff:1981um,Chapline:1982ww}, up to total derivatives, 
\be
\ba{l}
\displaystyle{
\int -2e^{-2d}\brcGo}\\
\displaystyle{\!\!
=\!\!\int\!\!\!\sqrt{-g}e^{-2\phi}\big(R+\!4\partial_{\mu}\phi\partial^{\mu}\phi-\!\textstyle{\frac{1}{12}}\tilde{H}_{\lambda\mu\nu}\tilde{H}^{\lambda\mu\nu\!}-\!\quarter F_{\mu\nu\dC}F^{\mu\nu\dC}
\big)},
\ea
\ee
where 
\be
\tilde{H}_{\lambda\mu\nu}=3\partial_{[\lambda}B_{\mu\nu]}-\omega_{\lambda\mu\nu}\,,
\ee 
which is invariant under the Yang--Mills gauge symmetry~(\ref{SYMMETRY}), or  specifically in components,
\[
\ba{ll}
\delta W_{\mu}{}^{\dA}=\partial_{\mu}\Lambda^{\dA}+f^{\dA}{}_{\dB\dC}W_{\mu}{}^{\dB}\Lambda^{\dC}\,,\quad&\quad
\delta B_{\mu\nu}=W_{[\mu}{}^{\dA}\partial_{\nu]}\Lambda_{\dA}\,.
\ea
\] 
However, we have not pinned down the gauge group to be either $\mathbf{E}_{8}\times \mathbf{E}_{8}$ or $\mathbf{SO}(32)$.  
Further, as the starting scalar curvatures~(\ref{cGobrcGo}) are at most two-derivative,  our final  result~(\ref{MAIN})  lacks the  higher-derivative gravitational Chern--Simons term~\cite{Green:1984sg}, \textit{c.f.~}\cite{Bedoya:2014pma,Coimbra:2014qaa,Lee:2015kba}.\\

\section{Reduction to heterotic string}
\vspace{5pt}
\noindent  Henceforth we discuss the worldsheet aspect of the DFT background~(\ref{ansatz}) involving   the maximally  non-Riemannian internal space. The  analysis of  the string  propagating on the most  general  non-Riemannian background was carried out  in \cite{Morand:2017fnv}, with the conclusion  that string becomes generically  (anti-)chiral.    To review this and  apply it to the current   non-Riemannian background~(\ref{ansatz}),  we recall the  doubled-yet-gauged string action~\cite{Hull:2006va},\,\cite{Lee:2013hma} (\textit{c.f.~}\cite{Park:2016sbw,Arvanitakis:2017hwb,Arvanitakis:2018hfn}),
\be
S_{\scriptstyle{\rm{string}}}={\textstyle{\frac{1}{4\pi\alpha^{\prime}}}}{\displaystyle{\int}}\rmd^{2}\sigma~\cL_{\scriptstyle{\rm{string}}}\,,
\label{stringA}
\ee
which contains a generic DFT-metric, 
\[
\cL_{\scriptstyle{\rm{string}}}=-\half\sqrt{-h}h^{\alpha\beta}\rmD_{\alpha}z^{\hA}\rmD_{\beta}z^{\hB}\hcH_{\hA\hB}
-\epsilon^{\alpha\beta}\rmD_{\alpha}z^{\hA}\cA_{\beta\hA}\,.
\label{stringaction}
\]
With  ambient doubled coordinates,  $z^{\hA}=(\tilde{z}_{\hat{\mu}},z^{\hat{\nu}})$, and an auxiliary gauge potential, $\cA_{\alpha}{}^{\hA}$,  a covariant derivative is introduced, 
\be
\ba{ll}
{\rmD_{\alpha} z^{\hA}:=\partial_{\alpha}z^{\hA}-\cA_{\alpha}{}^{\hA}}\,,\qquad&\quad
\cA_{\alpha}{}^{\hA}\partial_{\hA}=0\,.
\ea
\label{rmDz}
\ee
While this action is completely covariant with respect to the desired symmetries like      $\ODD$ T-duality, Weyl symmetry, and  worldsheet as well as  target-spacetime diffeomorphisms, it also     realizes  concretely   the assertion  that  doubled coordinates in DFT are actually gauged and each gauge orbit in the doubled coordinate system corresponds to a single physical point~\cite{Park:2013mpa}.  Specifically,  with the choice of the section, $\partial_{\hA}=(\tilde{\partial}^{\hat{\mu}},\partial_{\hat{\nu}})\equiv(0,\partial_{\hat{\nu}})$,  the condition on the auxiliary gauge potential~(\ref{rmDz})  is  solved by $\cA_{\alpha}{}^{\hA}=(\cA_{\alpha\hat{\mu}},0)$. Therefore,    half of the doubled coordinates, namely  the  tilde coordinates, are gauged,
\[
{\rmD_{\alpha} z^{\hA}=\big(\partial_{\alpha}\tilde{z}_{\hat{\mu}}-\cA_{\alpha\hat{\mu}}\,,\,\partial_{\alpha} z^{\hat{\nu}}\big)}\,.
\]
Upon the Riemannian background~(\ref{Riemann}),  the auxiliary gauge potential  appears quadratically in  the action~(\ref{stringA}). Then after integrating it out,   one  recovers   the standard undoubled string action with   $g_{\mu\nu}$ and $B_{\mu\nu}$.  On the  other hand, upon a generic   non-Riemannian DFT background which is characterized by two non-negative integers, $(n,\brn)$, the  components of the  auxiliary field,  $\cA_{\alpha}{}^{\hA}$,  appear linearly for  $n$ and $\brn$ directions, playing    the role of Lagrange multipliers. Consequently,   string becomes  chiral and anti-chiral over the $n$ and $\brn$ directions respectively~\cite{Morand:2017fnv}.

Now, for  the present   non-Riemannian background~(\ref{ansatz}),  in accordance with the decomposition of  the  ambient space~(\ref{split}),  we put  $y^{\Apr}=(\ty_{\mu^{\prime}},y^{\nu^{\prime}})$  for the internal doubled coordinates,   $x^{A}=(\tx_{\mu},x^{\nu})$    for the external ones,  and   $W_{\mu}{}^{\Apr}=(W_{\mu\nu^{\prime}},\tilde{W}_{\mu}{}^{\nu^{\prime}})$  for the gravivector. Further,   for the ambient  gauge potential,    $\cA_{\alpha}{}^{\hA}=(\cA_{\alpha}{}^{\Apr}, \cA_{\alpha}{}^{A})$, we set  the internal and the external ones as   $\cA_{\alpha}{}^{\Apr}=(A_{\alpha\mu^{\prime}},0)$ and $\cA_{\alpha}{}^{A}=(A_{\alpha\mu},0)$.  With this preparation and  (\ref{ansatz}),  (\ref{Riemann}),  the string action~(\ref{stringA}) becomes quadratic in the combination of the  gauge potential components,  $A_{\alpha\mu}-\tilde{W}_{\mu}{}^{\mu^{\prime}}A_{\alpha\mu^{\prime}}$,  and linear in $A_{\alpha\mu^{\prime}}$.  Hence,   the former leads to a Gaussian integral and  the latter plays the role of a Lagrange multiplier.  After all,   the   doubled-yet-gauged string action~(\ref{stringA})  reduces to 
$\frac{1}{2\pi\alpha^{\prime}}
\displaystyle{\int}\rmd^{2}\sigma~\cL_{\scriptstyle{\rm{Het}}}$,  with   
\be
\ba{lll}
\!\cL_{\scriptstyle{\rm{Het}}}\!&\!\!=\!\!\!&
-\half\sqrt{-h}h^{\alpha\beta}\partial_{\alpha}x^{\mu}\partial_{\beta}x^{\nu}
g_{\mu\nu}
+\half\epsilon^{\alpha\beta}\partial_{\alpha}x^{\mu}
\partial_{\beta}x^{\nu}B_{\mu\nu}\\
{}&{}&+\half\epsilon^{\alpha\beta}
(\partial_{\alpha}\ty_{\mu^{\prime}}+\partial_{\alpha}x^{\mu}W_{\mu\mu^{\prime}})
(\partial_{\beta}y^{\mu^{\prime}}+\partial_{\beta}x^{\lambda}\tilde{W}_{\lambda}{}^{\mu^{\prime}})\\
{}&{}&+\half\epsilon^{\alpha\beta}
(\partial_{\alpha}\tx_{\mu}-W_{\mu\Apr}\partial_{\alpha}y^{\Apr})\partial_{\beta}x^{\mu}\,.
\ea
\label{LHet}
\ee
Furthermore, $\partial_{\alpha}y^{\mu^{\prime}}+\partial_{\alpha}x^{\mu}\tilde{W}_{\mu}{}^{\mu^{\prime}}$ must  be \textit{chiral}:
\be
\left(h^{\alpha\beta}+\textstyle{\frac{1}{\sqrt{-h}}}\epsilon^{\alpha\beta}\right)\!\left(\partial_{\beta}y^{\mu^{\prime}}+\partial_{\beta}x^{\mu}\tilde{W}_{\mu}{}^{\mu^{\prime}}\right)=0\,.
\label{chiral}
\ee
Especially when the gravivector, $W_{\mu}{}^{\dA}$,  is trivial,   the second and  third lines in (\ref{LHet})  merge into  a   known  topological term, $\half\epsilon^{\alpha\beta}\partial_{\alpha}\tilde{z}_{\hat{\mu}}
\partial_{\beta}z^{\hat{\mu}}$~\cite{Giveon:1991jj,Hull:2006va},  while  (\ref{chiral})  gets  simplified to make    the internal coordinates  chiral: 
\be
y^{\mu^{\prime}}(\tau,\sigma)=y^{\mu^{\prime}}(0,\tau+\sigma)\,.
\ee
This implies -- at least classically~\cite{Morand:2017fnv,Ko:2015rha} --  that a closed string subject to a periodic boundary condition cannot vibrate  in the internal space.  In summary, upon the DFT background~(\ref{ansatz}) with the maximally non-Riemannian internal space and trivial gravivector,  the external  fields, $x^{\mu}$,  are described by the usual string action corresponding to the first line in (\ref{LHet}), and  consist  of the equal number of  left-moving and   right-moving sectors. On the other hand,      the internal  fields, $y^{\mu^{\prime}}$, become all chiral, which   agrees with the rigidity~(\ref{rigid}).\\

\section{Conclusion}
\vspace{5pt}
\noindent The maximally non-Riemannian DFT background specified by  the DFT-metric,  $\cH^{\prime}_{\Apr\Bpr}=\cJ^{\prime}_{\Apr\Bpr}$~(\ref{internal}),    is   singled out  to be  completely $\inODD$ symmetric and rigid: it does not admit any  linear fluctuation, $\delta\cH^{\prime}_{\Apr\Bpr}=0$~(\ref{rigid}), nor graviscalar moduli, and the coset structure is trivial~(\ref{coset}).  It is the most symmetric vacuum in DFT.  \\
\indent For  the  DFT Kaluza--Klein ansatz, (\ref{ansatz}) and (\ref{ansatzV}), we  set the internal space to be maximally non-Riemannian, performed   a  Scherk--Schwarz twist~(\ref{twist1}),~(\ref{twist2}), and  computed the ambient higher $(D^{\prime}+D)$-dimensional DFT scalar curvatures, $+\hcGo$ and  $-\hbrcGo$~(\ref{cGobrcGo}), which lead to the   $\ODD$-manifest  formulation of the  non-Abelian heterotic supergravity~(\ref{MAIN}).  Only the former, $+\hcGo$,  contains a DFT-cosmological constant.  In this way,  \textit{supergravity and Yang-Mills theory can be unified into a higher-dimensional pure Stringy Gravity, free of  moduli.}  \\   
\indent Plugging   the \textit{same} non-Riemannian Kaluza--Klein ansatz~(\ref{ansatz}) into  the doubled-yet-gauged string action~(\ref{stringA})  may reproduce  the usual  heterotic  string action,  
\[{\resizebox{.97\hsize}{!}{\,${\half\int_{\Sigma}}
\big({-\sqrt{-h}h^{\alpha\beta}}
g_{\mu\nu}+\epsilon^{\alpha\beta}B_{\mu\nu}\big)\partial_{\alpha}x^{\mu}
\partial_{\beta}x^{\nu}+\epsilon^{\alpha\beta}\partial_{\alpha}\tilde{z}_{\hat{\mu}}
\partial_{\beta}z^{\hat{\mu}}\,,$}}
\] 
with chiral internal coordinates,  $\big(h^{\alpha\beta}+\frac{\epsilon^{\alpha\beta}}{\sqrt{-h}}\big)\partial_{\beta}y^{\mu^{\prime}}=0$.   This is indeed  the case when the Yang--Mills sector is trivial.  Therefore, \textit{heterotic string theory may have  a higher-dimensional origin with non-Riemannian internal space.}

Our analyses have been focused on the bosonic sectors, both on the target-spacetime and   on the world-sheet.  Inclusion of fermions is  highly  desired, \textit{c.f.~}\cite{Jeon:2011sq,Jeon:2012hp,Park:2016sbw}.   
We also leave the exploration of  the worldsheet aspect of the somewhat mysterious  `relaxed'  section condition designed  for the  Scherk--Schwarz twist~\cite{Geissbuhler:2011mx,Aldazabal:2011nj,Grana:2012rr,Dibitetto:2012rk,Cho:2015lha} for future work with  nontrivial gravivector, $W_{A}{}^{\Apr}\neq0$,  for the action~(\ref{LHet}).  
Uplift of the Standard Model of particle physics  coupled to DFT~\cite{Choi:2015bga} to higher dimensions (bottom-up) would be also of interest, as well as applications to  string compactifications  (top-down), possibly on other types of non-Riemannian internal  spaces~\cite{33}.\\


\noindent\textit{Note added.}  \\
Motivated by the \href{https://arxiv.org/abs/1808.10605}{first version of this work on arXiv},  Berman,  Blair, and Otsuki  explored  non-Riemannian geometries in M-theory~\cite{Berman:2019izh}. In particular, they pointed out that the maximally non-Riemannian $E_{8}$ background naturally  realizes the topological  (hence `moduli-free') phase  of  Exceptional Field Theory~\cite{Hohm:2018ybo}. \\

\noindent\textit{Acknowledgments.}\\  {We  wish to thank Stephen Angus,  Kanghoon Lee for useful comments,  and     Wonyoung Cho for  valuable help at the early stage of the project.  This work  was  supported by  the National Research Foundation of Korea   through  the Grants,  NRF-2016R1D1A1B01015196  and NRF-2018H1D3A1A01030137~(Korea Research Fellowship Program).}
\hfill


\begin{thebibliography}{99}
\bibitem{LR}
Equations (\ref{defV1})  and (\ref{defV2})  imply respectively     that    $\big(V_{A}{}^{p},\brV_{A}{}^{\brp}\big)$ is the right and  left inverse of $\big(V^{B}{}_{q},\brV^{B}{}_{\brq}\big)^{T}$. Since  left inverse  is the same as   right inverse in general,  (\ref{defV1})  and (\ref{defV2})  are equivalent.
\bibitem{Quetient}
The numerators in the quotient groups of (\ref{coset}) are not  target-spacetime  T-duality group:  they act on the local Lorentz vector indices and correspond to  the stabilizer of the defining  relation of the DFT-vielbeins~(\ref{defV2}).
\end{thebibliography}
\end{document}